\documentstyle[12pt,axodraw,epsfig,epsf]{article}
\renewcommand{\baselinestretch}{1.2}
\begin{document}
\begin{center}
\vspace {1cm} {\large\bf Dynamical symmetry breaking of lambda and
vee-type three-level systems on quantization of the field modes}\\
\vspace{.5cm}{\bf Mihir Ranjan Nath{\footnote[1]
{mrnath\_95@rediffmail.com}} and Surajit Sen{\footnote[2]
{ssen55@yahoo.com}}
\\Department of Physics\\
Guru Charan College\\ Silchar 788004, India \\
\vspace{.5cm}
Asoke Kumar Sen{\footnote[3]{asokesen@sancharnet.in}}\\
Department of Physics \\ Assam University\\
Silchar 788011, India\\
\vspace{.5cm}
 Gautam Gangopadhyay{\footnote[4]{gautam@bose.res.in}}\\
S N Bose National Centre for Basic Sciences\\
JD Block, Sector III
\\Salt Lake City, Kolkata 700098, India}\\
\end{center}
\vspace{.5cm}
\begin{abstract}
\par
We develop a scheme to construct the Hamiltonians of the lambda, vee
and cascade type of three-level configurations using the generators
of $SU(3)$ group. It turns out that this approach provides a well
defined selection rule to give different Hamitonians for each
configurations. The lambda and vee type configurations are exactly
solved with different initial conditions while taking the two-mode
classical and quantized fields . For the classical field, it is
shown that the Rabi oscillation of the lambda model is similar to
that of the vee model and the dynamics of the vee model can be
recovered from lambda model and vice versa simply by inversion. We
then proceed to solve the quantized version of both models
introducing a novel Euler matrix formalism. It is shown that this
dynamical symmetry exhibited in the Rabi oscillation of two
configurations for the semiclassical models is completely destroyed
on quantization of the field modes. The symmetry can be restored
within the quantized models when the field modes are both in the
coherent states with large average photon number which is depicted
through the collapse and revival of the Rabi oscillations.\\
\end{abstract}
\pagebreak
\renewcommand{\baselinestretch}{1.6}
\begin{center}
\large I.Introduction
\end{center}
\par
Quantum Optics gives birth to many novel proposals which are within
reach of present-day ingenious experiments performed with intense
narrow-band tunable laser and high-Q superconducting cavity [1].
Major thrust in the atomic, molecular and optical experiments
primarily involves the coherent manipulation of the quantum states
which may be useful to verify several interesting results of quantum
information theory and also the experimental realization of the
quantum computer [2,3].
The actual number of the quantum mechanical states of atoms involved
in the interaction with light is of much importance in these days
since many coherent effects are due to the level structure of the
atom. It is well-known that the two-level system and its quantized
version, namely, the Jaynes-Cummings model (JCM), have been proved
to be an useful theoretical laboratory to understand many subtle
issues of the cavity electrodynamics [4,5]. The two-level system is
modeled using the Pauli's spin matrices - the spin-half
representation of $SU(2)$ group, where apart from the level number,
the spectrum is designated by the photon number as the quantum
number. A natural but non-trivial extension of the JCM is the
three-level system and it exhibits plethora of optical phenomena
such as, two-photon coherence [6], resonance Raman scattering [7],
double resonance process [8],population trapping [9], three-level
super radiance [10], three-level echoes [11], STIRAP [12], quantum
jump [13], quantum zeno effect [14], Electromagnetically Induced
Transparency [15,16] etc. There are three distinct schemes of the
three-level configurations which are classified as the lambda, vee
and cascade systems respectively. The Hamiltonians of these
configurations are generally modeled by two two-level systems
coupled by the two modes of cavity fields of different frequencies
[17,18].
Although these Hamiltonians
succeed in revealing several phenomena [19,20], however, their $ad$
$hoc$ construction subsides the underlying symmetry and its role in
the population dynamics of these systems. The connection between the
$SU(N)$ symmetry and the $N$-level system in general, was
investigated extensively in recent past [21-27]. These studies not
only mimic the possible connection between quantum optics with the
octet symmetry, well-known paradigm of particle physics, but for
$N=3$, it also reveals several interesting results such as the
realization of the eight dimensional Bloch equation, existence of
non-linear constants [18,22], population transfer via continuum
[28], dynamical aspects of three-level system in absence of
dissipation [29] etc. However, inspite of these progress, a general
formalism as well as the $ab$ $initio$ solutions of all three
configurations
are yet to be developed for the reasons mentioned below.
\par
The generic model Hamiltonian of a three-level configuration with
three well-defined energy levels can be represented by the hermitian
matrix
\begin{flushleft}
\hspace{1.5in}
 $H   = \left[ {\begin{array}{*{20}c}
   \Delta_{3} & h_{32} & h_{31}  \\
   h_{32} & \Delta_2 & h_{21}  \\
   h_{31} & h_{21} & \Delta_1  \\
\end{array}} \right],\hfill (1)$
\end{flushleft}
where $h_{ij} (i,j=1,2,3)$ be the matrix element of specific
transition and $\Delta_i$ be the detuning which vanishes at
resonance. We note that from Eq.(1), the lambda system, which
corresponds to the transition $1\leftrightarrow 3\leftrightarrow2$
shown in Fig.1a, can be described by the Hamiltonian with elements
$h_{21}=0$, $h_{32}\neq0$ and $h_{31}\neq0$. Similarly the vee
model, characterized by the transition
$3\leftrightarrow1\leftrightarrow2$ shown in Fig.1b, corresponds to
the elements $h_{21}\neq0$, $h_{32}=0$ and $h_{31}\neq0$ and for the
cascade model we have transition $1\leftrightarrow 2
\leftrightarrow3 $, we have $h_{21}\neq0$, $h_{32}\neq0$ and
$h_{31}=0$ respectively. Thus we have distinct Hamiltonian for three
different configurations which can be read off from Eq.(1)
shown in Fig.1. This definition,
however, differs from the proposal advocated by Hioe and Eberly, who
argued the order of the energy levels to be $E_1<E_3<E_2$ for the
lambda system, $E_2<E_3<E_1$ for the vee system and $E_1<E_2<E_3$
for the cascade system respectively [18,21,22]. In their scheme, the
level-2 is always be the intermediary level which becomes the upper,
lower and middle level to generate the lambda, vee and cascade
configurations respectively. It is worth noting that, if we follow
their scheme, these energy conditions map all three three-level
configurations to a unique cascade Hamiltonian described by the
matrix with elements $h_{12}\neq0$, $h_{23}\neq0$ and $h_{13}=0$ in
Eq.(1).
Thus because of the similar structure of the model Hamiltonian, if
we start formulating the solutions of the lambda, vee
and cascade configurations, 
then it would led to same spectral feature. Furthermore, due to the
same reason, the eight dimensional Bloch equation always remains
same for all three models [18,22]. Both of these consequences go
against the usual notion because wide range of coherent phenomena
mentioned above arises essentially due to different class of the
three-level configurations. Thus it is worth pursing to formulate a
comprehensive approach, where we have distinct Hamiltonian for three
configurations without altering the second level for each model.
\par
The problem of preparing multilevel atoms using one or more laser
pulses is of considerable importance from experimental point of
view. Thus the completeness of the study of the three-level systems
requires the exact solution of these models to find the probability
amplitudes of all levels, the effect of the field quantization on
the population oscillation and, most importantly, the observation of
the collapse and revival effect. In recent past, the three-level
systems and its several ramifications were extensively covered
in a general framework of the $SU(N)$ group having $N$-levels
[21-27,30,31]. Also, the semiclassical model [24,32,33] and its
fully quantized version [23,34,35] are studied, but to our
knowledge, the pursuit of the exact solutions of different
three-level systems in the spirit of the theory of Electron Spin
Resonance (ESR) model and JCM, are still to be facilitated
analytically.
\par
In a recent paper, we have studied the exact solutions of the
equidistant cascade system interacting with the single mode
classical and quantized field with different initial conditions
[36]. It is shown that for the semiclassical model the Rabi
oscillation exhibits a symmetric pattern of evolution, which is
destroyed on quantization of the cavity field. We also show that
this symmetry
is restored by taking the cavity mode to be the coherent state
indicating the proximity of the coherent state to the classical
field. We have further studied the equidistant cascade four-level
system and obtain similar conclusions [37]. To extend above studies
for the lambda and vee models we note that the vee configuration can
be obtained from the lambda configuration simply by inversion.
However, it is worth noting that, the lambda configuration is
associated with processes such as STIRAP [12], EIT [15,16] etc,
while the vee configuration corresponds to the phenomena such as
quantum jump [13], quantum zeno effect [14], quantum beat [3] etc
indicating that both the processes are fundamentally different. It
is therefore natural to examine the inversion symmetry between the
models by comparing their Rabi oscillations and study the effect of
the field quantization on that symmetry. The comparison shows that
the inversion symmetry exhibited by the semiclassical models is
completely spoiled on quantization of the cavity modes indicating
the non-trivial role of the vacuum fluctuation in the symmetry
breaking.
\par
The remaining part of the paper is organized as follows. In
Section-II, we discuss the basic tenets of the $SU(3)$ group
necessary to develop the Hamiltonian of all possible three-level
configurations. Section-III deals with the solution of the lambda
model taking the two field modes as the classical fields and then in
Section IV we proceed to solve the corresponding quantized version
of the model using a novel Euler matrix formalism. Section-V and VI
we present similar calculation for the vee model taking the mode
fields to be first classical and then quantized respectively. In
Section-VII we compare the population dynamics in both models
and discuss its implications. Finally in Section-VIII we
conclude our results.
\par
\begin{center}
\large II.The Models
\end{center}
\par
The most general Hamiltonian of a typical three-level configuration
is given by Eq.(1) which contains several non-zero matrix elements
showing all possible allowed transitions. To show how the $SU(3)$
symmetry group provides a definite scheme of selection rule which
forbids any one of the three transitions to give the Hamiltonian of
a specific model, let us briefly recall the tenets of $SU(3)$ group
described by the Gell-Mann matrices, namely,
\begin{flushleft}
\hspace{0.5in}
 $\lambda_1   = \left[ {\begin{array}{*{20}c}
   0 & 1 & 0  \\
   1 & 0 & 0  \\
   0 & 0 & 0  \\
\end{array}} \right],
\hspace{0.35in} \lambda_ 2   = \left[ {\begin{array}{*{20}c}
   0 & -i & 0  \\
   i & 0 & 0  \\
   0 & 0 & 0  \\
\end{array}} \right],
\hspace{0.35in} \lambda_3 = \left[ {\begin{array}{*{20}c}
   1 & 0 & 0  \\
   0 & -1 & 0  \\
   0 & 0 & 0  \\
\end{array}} \right],$
\end{flushleft}

\begin{flushleft}
\hspace{0.5in}
 $\lambda_4   = \left[ {\begin{array}{*{20}c}
   0 & 0 & 1  \\
   0 & 0 & 0  \\
   1 & 0 & 0  \\
\end{array}} \right],
\hspace{0.35in} \lambda_5   = \left[ {\begin{array}{*{20}c}
   0 & 0 & -i  \\
   0 & 0 & 0  \\
   i & 0 & 0  \\
\end{array}} \right],
\hspace{0.35in} \lambda_6   = \left[ {\begin{array}{*{20}c}
   0 & 0 & 0  \\
   0 & 0 & 1  \\
   0 & 1 & 0  \\
\end{array}} \right],$
\end{flushleft}

\begin{flushleft}
\hspace{0.5in}
 $\lambda_7   = \left[ {\begin{array}{*{20}c}
   0 & 0 & 0  \\
   0 & 0 & -i  \\
   0 & i & 0  \\
\end{array}} \right],
\hspace{0.35in} \lambda_8   = \frac{1}{\sqrt{3}}\left[
{\begin{array}{*{20}c}
   1 & 0 & 0  \\
   0 & 1 & 0  \\
   0 & 0 & -2  \\
\end{array}} \right].\hfill (2)$
\end{flushleft}
These matrices follow the following commutation and
anti-commutation relations
\begin{flushleft}
\hspace{0.5in} $[\lambda_{i},\lambda_{j}]=2 i f_{ijk}\lambda_k$,
\hspace{0.35in}
$\{\lambda_{i},\lambda_{j}\}=\frac{4}{3}\delta_{ij}+2
d_{ijk}\lambda_k$,\hfill (3)
\end{flushleft}
respectively, where $d_{ijk}$ and $f_{ijk}$ ($i,j=1,2,..8$)
represent completely symmetric and completely antisymmetric
structure constants which characterizes $SU(3)$ group [39]. It is
customary to define the shift operators $T$, $U$ and $V$ spin as
\begin{flushleft}
\hspace{0.5in} $T_{\pm}=\frac{1}{2}(\lambda_1 \pm i\lambda_2)$,
\hspace{0.35in} $U_{\pm}=\frac{1}{2}(\lambda_{6} \pm
i\lambda_{7})$, \hspace{0.35in} $V_{\pm}=\frac{1}{2}(\lambda_{4}
\pm i \lambda_5)$. \hfill (4)
\end{flushleft}
They satisfy the closed algebra
\begin{flushleft}
\hspace{0.5in} $[U_{+},U_{-}]=U_3$,\hspace{0.35in}
$[V_{+},V_{-}]=V_3$, \hspace{0.35in} $[T_{+},T_{-}]=T_3$,
\hfill(5)\\
\hspace{0.5in} $[T_{3},T_{\pm}]=\pm 2T_\pm$, \hspace{0.25in}
$[T_{3},U_{\pm}]=\mp U_\pm$,\hspace{0.22in}
$[T_{3},V_{\pm}]=\pm V_\pm$,\\

\hspace{0.5in} $[V_{3},T_{\pm}]=\pm T_\pm$, \hspace{0.25in}
$[V_{3},U_{\pm}]=\pm U_\pm$,\hspace{0.22in}
$[V_{3},V_{\pm}]=\pm 2V_\pm$,\\

\hspace{0.5in} $[U_{3},T_{\pm}]=\mp T_\pm$, \hspace{0.25in}
$[U_{3},U_{\pm}]=\pm 2U_\pm$,\hspace{0.22in}
$[U_{3},V_{\pm}]=\pm V_\pm$,\\

\hspace{0.5in} $[T_{+},V_{-}]=-U_-$,\hspace{0.32in}
$[T_{+},U_{+}]=V_+$, \hspace{0.38in} $[U_{+},V_{-}]=T_-$, \\

\hspace{0.5in} $[T_{-},V_{+}]=U_+$,\hspace{0.32in}
$[T_{-},U_{-}]=-V_-$, \hspace{0.38in} $[U_{-},V_{+}]=-T_+$, \\
\end{flushleft}
where the diagonal terms are $T_3=\lambda_3$,
$U_3=(\sqrt{3}\lambda_8-\lambda_3)/2$ and
$V_3=(\sqrt{3}\lambda_8+\lambda_3)/2$, respectively.
\par
The Hamiltonian of the semiclassical lambda model is given by
\begin{flushleft}
\hspace{0.5in} ${\rm H^{\Lambda}} = {\rm H_I^{\Lambda}}+{\rm
H_{II}^{\Lambda}}$, \hfill(6a)
\end{flushleft}
where the unperturbed and interaction parts including the detuning
terms are given by
\begin{flushleft}
\hspace{0.5in} ${\rm H_I^{\Lambda}} = \hbar(\Omega
_1-\omega_1-\omega_2) V_3+\hbar(\Omega _2-\omega_1-\omega_2) {\rm
T}_3$, \hfill(6b)
\end{flushleft}
and
\begin{flushleft}
\hspace{0.5in} ${\rm
H_{II}^{\Lambda}}=\hbar(\Delta^{\Lambda}_1V_3+\Delta^{\Lambda}_2T_3)
+$
\end{flushleft}
\begin{flushleft}
\hspace{0.5in} $\hbar \kappa _1 (V_ + \exp ( - i\Omega _1 t) + V_
- \exp (i\Omega _1 t))+ \hbar \kappa _2 (T_+\exp(-i\Omega_2 t)+
T_-\exp (i\Omega_2 t))$, \hfill(6c)
\end{flushleft}
respectively. In Eq.(6), $\Omega_i$ ($i=1,2$) are the external
frequencies of the bi-chromatic field, $\kappa_{i}$ are the coupling
parameters and $\hbar \omega _1 (=-{\rm E}_1), \hbar \omega _2 (=
-{\rm E}_2), \hbar (\omega _2 + \omega _1 )({=\rm E}_3)$ be the
respective energies of the three levels.
$\Delta^{\Lambda}_1=(2\omega_1+\omega_2-\Omega_1)$ and
$\Delta^{\Lambda}_2=(\omega_1+2\omega_2-\Omega_2)$ represent the
respective detuning from the bi-chromatic external frequencies as
shown in Fig.1.
\par
Proceeding in the same way, the semiclassical vee type three-level
system can be written as
\begin{flushleft}
\hspace{0.5in} ${\rm H^{V}} ={\rm  H_I^{V}}+{\rm H_{II}^{V}},$
\hfill(7a)
\end{flushleft}
where 
\begin{flushleft}
\hspace{0.5in} ${\rm H_I^{V}} = \hbar(\Omega _1-\omega_1-\omega_2)
V_3+\hbar(\Omega _2-\omega_1-\omega_2) {\rm U}_3,$ \hfill(7b)
\end{flushleft}
and
\begin{flushleft}
\hspace{0.5in} ${\rm
H_{II}^{V}}=\hbar(\Delta^{V}_1V_3+\Delta^{V}_2U_3) +$
\end{flushleft}
\begin{flushleft}
\hspace{0.5in} $\hbar \kappa _1 (V_ + \exp ( - i\Omega _1 t) + V_
- \exp (i\Omega _1 t))+ \hbar \kappa _2 (U_+\exp(-i\Omega_2 t)+
U_-\exp (i\Omega_2 t))$ \hfill(7c)
\end{flushleft}
where $\Delta^V_1=(2\omega_1+\omega_2-\Omega_1)$ and
$\Delta^V_2=(2\omega_2+\omega_1-\Omega_2)$ be the detuning shown in
Fig.2.
\par
Similarly the semiclassical cascade three-level model is given by
\begin{flushleft}
\hspace{0.5in} ${\rm H^{\Xi}} = {\rm H_I^{\Xi}}+{\rm
H_{II}^{\Xi}}$, \hfill(8a)
\end{flushleft}
where
\begin{flushleft}
\hspace{0.5in} ${\rm H_I^{\Xi}} = \hbar(\Omega _1+\omega_2-\omega_1)
{\rm U}_3+\hbar(\Omega _2+\omega_1-\omega_2) {\rm T}_3$, \hfill(8b)
\end{flushleft}
and
\begin{flushleft}
\hspace{0.5in} ${\rm
H_{II}^{\Xi}}=\hbar(\Delta_1^{\Xi}U_3+\Delta_2^{\Xi}T_3)+$
\end{flushleft}
\begin{flushleft}
\hspace{0.5in} $\hbar \kappa _1 (U_ + \exp ( - i\Omega _1 t) + U_
- \exp (i\Omega _1 t))+ \hbar \kappa _2 (T_+\exp(-i\Omega_2 t)+
T_-\exp (i\Omega_2 t))$ \hfill(8c)
\end{flushleft}
respectively with respective detuning
$\Delta^{\Xi}_1=(2\omega_1-\omega_2-\Omega_1)$ and
$\Delta^{\Xi}_2=(2\omega_2-\omega_1-\Omega_2)$.
\par
Taking the fields to be the quantized cavity fields, in the rotating
wave approximation, the Hamiltonian of the quantized lambda
configuration is given by
\begin{flushleft}
\hspace{0.5in} ${\it H^{\Lambda}} = H_I^{\Lambda}+H_{II}^{\Lambda}$,
\hfill(9a)
\end{flushleft}
where,
\begin{flushleft}
\hspace{0.5in} ${\it H_I^{\Lambda}} = \hbar(\Omega _2 - \omega_1 -
\omega_2) {\rm T}_3 + \hbar (\Omega _1 - \omega_1 - \omega_2) V_3 +
\sum\limits_{j = 1}^2\Omega_j{a_j^{\dag} a_j}$, \hfill(9b)
\end{flushleft}
\begin{flushleft}
\hspace{0.5in} ${\it H_{II}^{\Lambda}}=\hbar\Delta_1^{\Lambda} V_3 +
\hbar\Delta_2^{\Lambda} {\rm T}_3+  \hbar g_{1} (V_{+} a_1 + V_{-}
a_1^{\dagger}) + \hbar g_{2} (T_{+}a_2 + T_{-}a_2^ {\dagger}
),$\hfill(9c)
\end{flushleft}
where $a^\dagger_i$ and $a_i$ ($i=1,2$) be the creation and
annihilation operators of the cavity modes, $g_i$ be the coupling
constants and $\Omega_i$ be the mode frequencies. Proceeding in the
similar pattern, the Hamiltonian of the quantized vee system is
given by
\begin{flushleft}
\hspace{0.5in} ${\it H^{V}}=H_I^{V}+H_{II}^{V}$, \hfill(10a)
\end{flushleft}
where,
\begin{flushleft}
\hspace{0.5in} ${\it H_I^{V}} = \hbar(\Omega _2 - \omega_1 -
\omega_2) {\rm U}_3 + \hbar (\Omega _1 - \omega_1 - \omega_2) V_3 +
\sum\limits_{j = 1}^2\Omega_j{a_j^{\dag} a_j }$ \hfill(10b)
\end{flushleft}
\begin{flushleft}
\hspace{0.5in} ${\it H_{II}^{V}}=\hbar\Delta_1^{V} V_3 +
\hbar\Delta_2^{V} {\rm U}_3+  \hbar g_{1} (V_{+} a_1 + V_{-}
a_1^{\dagger}) + \hbar g_{2} (U_{+}a_2 + U_{-}a_2^ {\dagger}
),$\hfill(10c)
\end{flushleft}
respectively. Similarly the Hamiltonian of the quantized cascade
system reads
\begin{flushleft}
\hspace{0.5in} ${\it H^{\Xi}} = H_I^{\Xi}+H_{II}^{\Xi}$, \hfill(11a)
\end{flushleft}
where
\begin{flushleft}
\hspace{0.5in} ${\it H_I^{\Xi}} = \hbar(\Omega _2 - \omega_1 -
\omega_2) {\rm T}_3 + \hbar (\Omega _1 - \omega_1 - \omega_2) U_3 +
\sum\limits_{j = 1}^2\Omega_j{a_j^{\dag} a_j }$, \hfill(11b)
\end{flushleft}
\begin{flushleft}
\hspace{0.5in} ${\it H_{II}^{\Xi}}=\hbar\Delta_1^{\Xi} U_3 +
\hbar\Delta_2^{\Xi} {\rm T}_3+  \hbar g_{1} (U_{+} a_1 + U_{-}
a_1^{\dagger}) + \hbar g_{2} (T_{+}a_2 + T_{-}a_2^ {\dagger}
).$\hfill(11c)
\end{flushleft}
Using the algebra given in Eq.(5) and that of field operators, it is
easy to check that $[{\it H_I^i},{\it H_{II}^i}]=0$ for
$\Delta^i_1=-\Delta^i_2$ ($i=\Lambda$ and $V$) for the lambda and
vee model and $\Delta^\Xi_1=\Delta^\Xi_2$ for the cascade model
which are identified as the two photon resonance condition and equal
detuning conditions, respectively [18,21,22,24,26]. This ensures
that each piece of the Hamiltonian has the simultaneous eigen
functions. Thus we note that, unlike Ref.[18,21,22], precise
formulation of the aforementioned three-level configurations require
the use of a subset of Gell-Mann $\lambda_i$ matrices rather than
the use of all matrices. We now proceed to solve the lambda and vee
configurations for the classical and the quantized field separately.
\begin{center}
\large III.The semiclassical lambda system
\end{center}
\par
At zero detuning the Hamiltonian of the lambda type three-level
system is given by
\begin{flushleft}
\hspace{1in}
 $H^{\Lambda}   = \left[ {\begin{array}{*{20}c}
   \hbar(\omega_1+\omega_2) & \hbar\kappa_2 \exp[-i\Omega_2 t] &
 \hbar\kappa_1 \exp[-i\Omega_1 t]    \\
   \hbar \kappa_2 \exp[i\Omega_2 t] & -\hbar\omega_2 & 0  \\
   \hbar\kappa_1 \exp[i\Omega_1 t] & 0 & -\hbar\omega_1  \\
\end{array}} \right].\hfill (12)$
\end{flushleft}
The solution of the Schrodinger equation corresponding to
Hamiltonian (12) is given by
\begin{flushleft}
\hspace{1in}$ \Psi (t) = C_1(t) \left| 1 \right\rangle  + C_2(t)
\left| 2 \right\rangle  + C_3(t) \left| 3 \right\rangle$\hfill(13)
\end{flushleft}
where $C_1(t)$, $C_2(t)$ and $C_3(t)$ be the time-dependent
normalized amplitudes of the lower, middle and upper levels with the
respective basis states,
\begin{flushleft}
\hspace{1in}$\left| 1 \right\rangle  = \left[ {\begin{array}{*{20}c}
   0  \\
   0  \\
   1  \\
\end{array}} \right],$
\hspace{.5in}$\left| 2 \right\rangle  = \left[
{\begin{array}{*{20}c}
   0  \\
   1  \\
   0  \\
\end{array}} \right],$
\hspace{0.5in}$\left| 3 \right\rangle  = \left[
{\begin{array}{*{20}c}
   1  \\
   0  \\
   0  \\
\end{array}} \right],$\hfill(14)
\end{flushleft}
respectively. We now proceed to calculate the probability amplitudes
of the three states. Substituting Eq.(13) in Schr\"{o}dinger
equation and equating the coefficients of $\left| 2 \right\rangle$ ,
$\left| 3 \right\rangle$ and $ \left| 1 \right\rangle$ from both
sides we obtain

\begin{flushleft}
\hspace{1in} $i\frac{{\partial C_3 }}{{\partial t}} = (\omega
_2+\omega_1)C_3 + \kappa _1 \exp ( - i\Omega _1 t)C_1 + \kappa _2
\exp ( - i\Omega _2 t)C_2,$\hfill(15a)
\end{flushleft}

\begin{flushleft}
\hspace{1in} $i\frac{{\partial C_2 }}{{\partial t}} =  - \omega _2
C_2 + \kappa _2 \exp (i\Omega _2 t)C_3,$\hfill(15b)
\end{flushleft}

\begin{flushleft}
\hspace{1in} $i\frac{{\partial C_1 }}{{\partial t}} =  - \omega _1
C_1 + \kappa _1 \exp (i\Omega _1 t)C_3.$\hfill(15c)
\end{flushleft}
\par
Let the solutions of Eqs.(15a-c) are of the following form,
\begin{flushleft}
\hspace{1.5in} $C_1  = {\rm A}_1 \exp (iS_1 t),$\hfill(16a)
\end{flushleft}
\begin{flushleft}
\hspace{1.5in} $C_2  = {\rm A}_2 \exp (iS_2 t),$\hfill(16b)
\end{flushleft}
\begin{flushleft}
\hspace{1.5in} $C_3  = {\rm A}_3 \exp (iS_3 t),$\hfill(16c)
\end{flushleft}
where $A_i$s' are the time independent constants to be determined.
Putting Eqs.(16a-c) in Eqs.(15a-c) we obtain
\begin{flushleft}
\hspace{1.5in} $(S_3  + \omega _2 + \omega_1 ){\rm A}_3  + \kappa _2
{\rm A}_2 + \kappa _1 {\rm A}_1  = 0$,\hfill(17a)
\end{flushleft}
\begin{flushleft}
\hspace{1.5in} $(S_3  + \Omega _2  - \omega _2 ){\rm A}_2  + \kappa
_2 {\rm A}_3 = 0$,\hfill(17b)
\end{flushleft}
\begin{flushleft}
\hspace{1.5in} $(S_3  + \Omega _1  - \omega _1 ){\rm A}_1  + \kappa
_1 {\rm A}_3 = 0$.\hfill(17c)
\end{flushleft}
In deriving Eqs.(17), the time independence of the amplitudes ${\rm
A}_3$, ${\rm A}_2$ and ${\rm A}_1$ are ensured by invoking the
conditions $S_2  = S_3  + \Omega _2$ and $S_1  = S_3 + \Omega _1$.
At resonance, we have $\Delta^{\Lambda}_1=0=-\Delta^{\Lambda}_2$
i.e, $(2\omega _2 + \omega _1 ) - \Omega _2 = 0=(\omega _2  +2
\omega _1 ) - \Omega _1$ and the solution of Eq.(17) yields

\begin{flushleft}
\hspace{1.5in} $S_3  =  - (\omega _2 + \omega_1)  \pm
\Delta$,\hfill(18a)
\end{flushleft}
\begin{flushleft}
\hspace{1.5in} $S_3  =  - (\omega _2 + \omega_1)$\hfill(18b)
\end{flushleft}
where $\Delta  = \sqrt {\kappa _1^2  + \kappa _2^2 }$ and we have
three values of $S_2$ and $S_1$ namely
\begin{flushleft}
\hspace{1.5in} $S_2^1  = \omega _2 ,S_2^{2,3}  = \omega _2  \pm
\Delta$,\hfill(19a)
\end{flushleft}
\begin{flushleft}
\hspace{1.5in} $S_1^1  = \omega _1 ,S_1^{2,3}  = \omega _1  \pm
\Delta$.\hfill(19b)
\end{flushleft}
Using Eqs.(18) and (19), Eq.(16) can be written as
\begin{flushleft}
\hspace{0.5in} $C_3 (t) = {\rm A}_3^1 \exp ( - i(\omega _2 +
\omega_1) t) $\end{flushleft}
\begin{flushleft}
\hspace{0.5in} $ + {\rm A}_3^2 \exp (i( - (\omega _2 + \omega_1) +
\Delta )t) + {\rm A}_3^3 \exp(i( - (\omega _2 + \omega_1)  - \Delta
)t)$,\hfill(20a)
\end{flushleft}
\begin{flushleft}
\hspace{0.5in} $C_2 (t) = {\rm A}_2^1 \exp (i\omega _2 t) + {\rm
A}_2^2 \exp (i(\omega _2  + \Delta )t) + {\rm A}_2^3 (i(\omega _2 -
\Delta )t)$,\hfill(20b)
\end{flushleft}
\begin{flushleft}
\hspace{0.5in} $C_1 (t) = {\rm A}_1^1 \exp (i\omega _1 t) + {\rm
A}_1^2 \exp (i(\omega _1  + \Delta )t) + {\rm A}_1^3 (i(\omega _1 -
\Delta )t)$,\hfill(20c)
\end{flushleft}
where $A_i$-s are the constants which can be calculated from the
 following
initial conditions:
\par
Case-I: At $t=0$ let the atom is in level-1, i.e. $C_1 (0)=1$, $C_2
(0)=0$, $C_3 (0)=0$. Using Eqns (15) and (20), the corresponding
time-dependent probabilities of the three levels are
\begin{flushleft}
\hspace{1.5in} $\left| {C_3 (t)} \right|^2  = \frac{{\kappa _1^2
}}{{\Delta ^2 }}\sin ^2 \Delta t$,\hfill(21a)
\end{flushleft}
\begin{flushleft}
\hspace{1.5in} $\left| {C_2 (t)} \right|^2  = 4\frac{{\kappa _1^2
\kappa _2^2 }}{{\Delta ^4 }}\sin ^4 \Delta t/2$,\hfill(21b)
\end{flushleft}
\begin{flushleft}
\hspace{1.5in} $\left| {C_1 (t)} \right|^2  = \frac{1}{{\Delta ^4
}}(\kappa _2^2 + \kappa _1^2 \cos \Delta t)^2$.\hfill(21c)
\end{flushleft}
\par
Case-II: If the atom is initially in level-2, i.e. $C_1 (0)=0$, $C_2
(0)=1$ and $C_3(0)=0$, the probabilities of the three states are
\begin{flushleft}
\hspace{1.5in} $\left| {C_3 (t)} \right|^2  = \frac{{\kappa _2^2
}}{{\Delta ^2 }}\sin ^2 \Delta t$,\hfill(22a)
\end{flushleft}
\begin{flushleft}
\hspace{1.5in} $\left| {C_2 (t)} \right|^2  = \frac{1}{{\Delta ^4
}}(\kappa _1^2 + \kappa _2^2 \cos \Delta t)^2$,\hfill(22b)
\end{flushleft}
\begin{flushleft}
\hspace{1.5in} $\left| {C_1 (t)} \right|^2  = 4\frac{{\kappa _1^2
\kappa _2^2 }}{{\Delta ^4 }}\sin ^4 \Delta t/2$.\hfill(22c)
\end{flushleft}
\par
Case-III: When the atom is initially in level-3, i.e. $C_1(0)=0$,
$C_2(0)=0$ and $C_3(0)=1$, the time evolution of the probabilities
of the three states are
\begin{flushleft}
\hspace{1.5in} $\left| {C_3 (t)} \right|^2  = \cos ^2 \Delta
t$,\hfill(23a)
\end{flushleft}
\begin{flushleft}
\hspace{1.5in} $\left| {C_2 (t)} \right|^2  = \frac{{\kappa _2^2
}}{{\Delta ^2 }}\sin ^2 \Delta t$,\hfill(23b)
\end{flushleft}
\begin{flushleft}
\hspace{1.5in} $\left| {C_1 (t)} \right|^2  = \frac{{\kappa _1^2
}}{{\Delta ^2 }}\sin ^2 \Delta t$.\hfill(23c) \\

We now proceed to solve the quantized version of the above model.
\end{flushleft}

\begin{center}
\large IV. The quantized lambda system
\end{center}
\par
We now consider the three-level lambda system interacting with a
bi-chromatic quantized fields described by the Hamiltonian Eq.(9).
At zero detuning the solution of the Hamiltonian is given by
\begin{flushleft}
\hspace{0.15in} $\left| {\Psi _{\Lambda} (t)} \right\rangle  =
\sum\limits_{n,m = 0}^\infty  {[C_1^{n-1,m + 1} (t)\left|{n-1,m +
1,1} \right\rangle+C_2^{n,m} (t)\left| {n,m,2} \right\rangle}
+C_3^{n - 1,m} (t)\left| {n -1,m,3} \right\rangle ],$
\end{flushleft}
\hspace {6in}(24)\\ where $n$ and $m$ represent the photon number
corresponding to two modes of the bi-chromatic fields. This
interaction Hamiltonian that couples the atom-field states ${\left|
{n - 1,m,3} \right\rangle }$, ${\left| {n,m,2} \right\rangle }$ and
${\left| {n-1,m+1,1} \right\rangle }$ and forms the lambda
configuration shown in Fig.1 is given by
\begin{flushleft}
\hspace{1.5in} ${\it H_{II}^{\Lambda}} = \hbar \left[
{\begin{array}{*{20}c}
   0 & {g_{2} \sqrt {n } } & {g_{1} \sqrt {m + 1} }  \\
   {g_{2} \sqrt {n } } & 0 & 0  \\
   {g_{1} \sqrt {m + 1} } & 0 & 0  \\
\end{array}} \right].$\hfill(25)
\end{flushleft}
The eigenvalues of the Hamiltonian are given by $\lambda _ {\pm}
=\pm \hbar \sqrt {n g_{2}^2+ (m+1)g_{1}^2}\quad (= \pm \hbar\Omega
_{nm})$ and $\lambda _0 = 0 ( = \Omega _{0})$, respectively with the
corresponding dressed eigenstates
\begin{flushleft}
\hspace{1.5in} $\left[ {\begin{array}{*{20}c}
   {\left| {nm, 3 } \right\rangle }  \\
   {\left| {nm,2} \right\rangle }  \\
   {\left| {nm, 1 } \right\rangle }  \\
\end{array}} \right] = T_{n,m}(g_1,g_2)\left[ {\begin{array}{*{20}c}
   {\left| {n-1,m,3} \right\rangle }  \\
   {\left| {n,m,2} \right\rangle }  \\
   {\left| {n-1,m+1,1} \right\rangle }  \\
\end{array}} \right].$\hfill(26)
\end{flushleft}
In Eq.(26), the dressed states are constructed by rotating the bare
states with the Euler matrix given by
\begin{flushleft}
\hspace{1.0in} $T_{n,m}(g_1,g_2) = \left[ {\begin{array}{*{20}c}
   {c_3c_2-c_1s_2s_3} & {c_3s_2-c_1c_2s_3} & {s_3s_1}  \\
   {-s_3c_2-c_1s_2c_3} & {-s_3s_2+c_1c_2c_3} & {c_3s_1}  \\
   {s_1s_2} & {-s_1c_2} & {c_1}  \\
\end{array}} \right]$\hfill(27)
\end{flushleft}
where $s_i=\sin\theta_i$ and $c_i=\cos\theta_i$ ($i=1,2,3$). The
elements of the matrix are found to
\begin{flushleft}
$T_{n,m}(g_1,g_2) = \left[ {\begin{array}{*{20}c}
   {\frac{1}{{\sqrt 2 }}} & {g_2 \sqrt {\frac{{n }}{{2(n g_{2}^2+
 (m+1)g_{1}^2)}}}} & {g_{1} \sqrt {\frac{{m +
1}}{{2(n g_{2}^2+ (m+1)g_{1}^2)}}}}  \\
   {0} & {g_{1} \sqrt {\frac{{m +
1}}{{n g_{2}^2+ (m+1)g_{1}^2}}}} & {-g_{2} \sqrt {\frac{{n }}{{n
 g_{2}^2+ (m+1)g_{1}^2}}}}  \\
   {-\frac{1}{{\sqrt 2 }}} & {g_{2} \sqrt {\frac{{n }}{{2(n g_{2}^2+
 (m+1)g_{1}^2)}}}} & {g_{1} \sqrt
{\frac{{m + 1}}{{2(n g_{2}^2+ (m+1)g_{1}^2)}}}}  \\
\end{array}} \right],$\hfill(28)
\end{flushleft}
with corresponding Euler angles,
\begin{flushleft}
$\theta_1=\arccos [\frac{\sqrt{1+m}g_1}{\sqrt{2(1+m)g_1^2+2 n
g_2^2}} ]$,
$\theta_2=-\arccos[-\frac{\sqrt{n}g_2}{\sqrt{(1+m)g_1^2+2n g_2^2}}
]$,
$\theta_3=\arccos[-\frac{\sqrt{2n}g_2}{\sqrt{(1+m)g_1^2+2n g_2^2}}
]$.
\end{flushleft}
\hspace {15.0cm}(29) \\
The time-dependent probability amplitudes of
the three levels are given by
\begin{flushleft}
\hspace{0.15in} $\left[ {\begin{array}{*{20}c}
   {C_3^{n - 1,m} (t)}  \\
   {C_2^{n,m} (t)}  \\
   {C_1^{n - 1,m + 1} (t)}  \\
\end{array}} \right] = T_{n,m}^{-1}(g_1,g_2) \left[
 {\begin{array}{*{20}c}
   {e^{- i\Omega _{nm} t} } & 0 & 0  \\
   0 & {e^{-i\Omega _{0} t}} & 0  \\
   0 & 0 & {e^{i\Omega _{nm} t} }  \\
\end{array}} \right]T_{n,m}(g_1,g_2)\left[ {\begin{array}{*{20}c}
   {C_3^{n - 1,m} (0)}  \\
   {C_2^{n,m} (0)}  \\
   {C_1^{n - 1,m + 1} (0)}  \\
\end{array}} \right].$
\end{flushleft}
\hspace {15.0cm}(30)\\
Now similar to the semiclassical model the probabilities
corresponding to different initial conditions are:
\par
Case-IV: When the atom is initially in level-1, i.e, $C_1^{n - 1,m +
1}=1$, $C_2^{n,m}=0$ and $C_3^{n-1,m}=0$, the time-dependent atomic
populations of the three states are given by
\begin{flushleft}
\hspace{1.5in} $\left| {C_3^{n-1,m} (t)} \right|^2  =
\frac{{(m+1)g_{1}^2}}{{\Omega_{nm}^2 }}\sin ^2 \Omega _{nm}
t$,\hfill(31a)
\end{flushleft}
\begin{flushleft}
\hspace{1.5in} $\left| {C_2^{n,m} (t)} \right|^2  =
4\frac{{g_{1}^2g_{2}^2n(m + 1)}}{{\Omega _{nm}^4 }}\sin ^4 \Omega
_{nm} t/2$,\hfill(31b)
\end{flushleft}
\begin{flushleft}
\hspace{1.5in} $\left| {C_1^{n - 1,m + 1} (t)} \right|^2  =
\frac{1}{{\Omega _{nm}^4 }}[n g_{2}^2+ (m+1)g_{1}^2\cos \Omega _{nm}
t]^2$.\hfill(31c)
\end{flushleft}
Case-V: When the atom is initially in level-2, i.e,
$C_1^{n-1,m+1}=0$, $C_2^{n,m}=1$ and $C_3^{n-1,m}=0$, the
probabilities of three states are
\begin{flushleft}
\hspace{1.5in} $\left| {C_3^{n - 1,m} (t)} \right|^2  = \frac{{n
g_{2}^2}}{{\Omega _{nm}^2 }}\sin ^2 \Omega _{nm} t$,\hfill(32a)
\end{flushleft}
\begin{flushleft}
\hspace{1.5in} $\left| {C_2^{n,m} (t)} \right|^2= \frac{1}{{\Omega
_{nm}^4 }}[(m+1)g_{1}^2+ n g_{2}^2\cos \Omega _{nm} t]^2$,\hfill(32b)
\end{flushleft}
\begin{flushleft}
\hspace{1.5in} $\left| {C_1^{n-1,m+1} (t)} \right|^2  = 4\frac{{
g_{1}^2g_{2}^2n(m+1)}}{{\Omega _{nm}^4 }}\sin ^4 \Omega _{mm}
t/2$.\hfill(32c)
\end{flushleft}
Case-VI: If the atom is initially in level-3, then we have $C_1^{n
-1,m +1}=0$, $C_2^{n,m}=0$ and $C_3^{n-1,m+1}=1$ and the
corresponding probabilities are
\begin{flushleft}
\hspace{1.5in} $\left| {C_3^{n-1,m} (t)} \right|^2=\cos ^2 \Omega
_{nm} t$,\hfill(33a)
\end{flushleft}
\begin{flushleft}
\hspace{1.5in} $\left| {C_2^{n,m} (t)} \right|^2  = \frac{{n g_{2}^2
}}{{\Omega _{nm}^2 }}\sin ^2 \Omega _{nm} t$,\hfill(33b)
\end{flushleft}
\begin{flushleft}
\hspace{1.5in} $\left| {C_1^{n - 1,m + 1} (t)} \right|^2  =
\frac{{(m+1)g_{1}^2}}{{\Omega _{nm}^2 }}\sin ^2 \Omega _{nm}
t$.\hfill(33c)
\end{flushleft}
We now proceed to evaluate the population oscillations of different
levels of the vee system with similar initial conditions. \pagebreak

\begin{center}
\large V.The semiclassical vee system
\end{center}
\par
At zero detuning, the Hamiltonian of the semiclassical three-level
vee system interacting with two-mode classical fields is given by
\begin{flushleft}
\hspace{1.5in}
 $H^V   = \left[ {\begin{array}{*{20}c}
   \hbar\omega_1 & 0 & \hbar\kappa_1 \exp[-i\Omega_1 t]    \\
   0 & \hbar\omega_2 & \hbar\kappa_2 \exp[-i\Omega_2 t]  \\
   \hbar\kappa_1 \exp[i\Omega_1 t] & \hbar \kappa_2 \exp[i\Omega_2 t]
   & -\hbar(\omega_1+\omega_2)  \\
\end{array}} \right].\hfill (34)$
\end{flushleft}
\par
Let the solution of the Schrodinger equation corresponding to
Eq.(34) is given by
\begin{flushleft}
\hspace{1.5in}$ \Psi (t) = C_1(t)\left| 1 \right\rangle  + C_2(t)
\left| 2 \right\rangle  + C_3(t)\left| 3 \right\rangle,$\hfill(35)
\end{flushleft}
where $C_1(t)$, $C_2(t)$ and $C_3(t)$ are the time-dependent
normalized amplitudes with the basis vectors defined in Eqs.(13). To
calculate the probability amplitudes of three states, substituting
Eq.(35) into the Schr\"{o}dinger equation we obtain
\begin{flushleft}
\hspace{0.5in} $i\frac{{\partial C_3 }}{{\partial t}} =  \omega _1
C_3 + \kappa _1 \exp (-i\Omega _1 t)C_1$,\hfill(36a)
\end{flushleft}

\begin{flushleft}
\hspace{0.5in} $i\frac{{\partial C_2 }}{{\partial t}} =   \omega
_2 C_2 + \kappa _2 \exp (-i\Omega _2 t)C_1$,\hfill(36b)
\end{flushleft}

\begin{flushleft}
\hspace{0.5in} $i\frac{{\partial C_1 }}{{\partial t}} =  -(\omega
_1+\omega_2)C_1 + \kappa _2 \exp ( i\Omega _2 t)C_2 + \kappa _1
\exp ( i\Omega _1 t)C_3$.\hfill(36c)
\end{flushleft}

Let the solutions of Eqs.(36) are of the following form:
\begin{flushleft}
\hspace{1.5in} $C_3(t)  = {\rm A}_3 \exp (iS_3 t),$\hfill(37a)
\end{flushleft}
\begin{flushleft}
\hspace{1.5in} $C_2(t)  = {\rm A}_2 \exp (iS_2 t),$\hfill(37b)
\end{flushleft}
\begin{flushleft}
\hspace{1.5in} $C_1(t) = {\rm A}_1 \exp (iS_1 t),$\hfill(37c)
\end{flushleft}
where $A_i$-s are the time independent constants to be determined from
 the
boundary conditions. From Eq.(36) and Eq.(37) we obtain
\begin{flushleft}
\hspace{1.5in} $(S_1  - \Omega _1  + \omega _1 ){\rm A}_3  +
\kappa _1 {\rm A}_1 = 0$,\hfill(38a)
\end{flushleft}
\begin{flushleft}
\hspace{1.5in} $(S_1  - \Omega _2  + \omega_2 ){\rm A}_2  + \kappa
_2 {\rm A}_1 = 0$,\hfill(38b)
\end{flushleft}
\begin{flushleft}
\hspace{1.5in} $(S_1  - \omega _2 - \omega_1 ){\rm A}_1  + \kappa
_2 {\rm A}_2 + \kappa _1 {\rm A}_3  = 0$.\hfill(38c)
\end{flushleft}
In deriving Eqs.(38), the time independence of the amplitudes ${\rm
A}_3$, ${\rm A}_2$ and ${\rm A}_1$ are ensured by invoking the
conditions $S_2  = S_1  - \Omega _2$ and $S_3  = S_1 - \Omega _1$.
At resonance, $\Delta^{V}_1=0=-\Delta^{V}_2$ i.e. $(2\omega _2 +
\omega _1 ) - \Omega _2 = 0=(\omega _2  +2 \omega _1 ) - \Omega _1$
and the solutions of Eq.(38) are given by
\begin{flushleft}
\hspace{1.5in} $S_1 = (\omega _1 + \omega_2)$\hfill(39a)
\end{flushleft}
\begin{flushleft}
\hspace{1.5in} $S_1 =  (\omega _1 + \omega_2)  \pm
\Delta$\hfill(39b)
\end{flushleft}
and we have three values of $S_2$ and $S_3$
\begin{flushleft}
\hspace{1.5in} $S_2^1  = -\omega _2 ,S_2^{2,3}  = -\omega _2  \pm
\Delta$\hfill(40a)
\end{flushleft}
\begin{flushleft}
\hspace{1.5in} $S_3^1  = -\omega _1 ,S_3^{2,3}  = -\omega _1  \pm
\Delta$.\hfill(40b)
\end{flushleft}
Using Eqs.(39) and (40), Eqs. (37) can be written as
\begin{flushleft}
\hspace{0.5in} $C_3 (t) = {\rm A}_3^1 \exp (-i\omega _1 t) + {\rm
A}_3^2 \exp (-i(\omega _1  + \Delta )t) + {\rm A}_3^3 (-i(\omega
_1 - \Delta )t)$,\hfill(41a)
\end{flushleft}

\begin{flushleft}
\hspace{0.5in} $C_2 (t) = {\rm A}_2^1 \exp (-i\omega _2 t) + {\rm
A}_2^2 \exp (-i(\omega _2  + \Delta )t) + {\rm A}_2^3 (-i(\omega
_2 - \Delta )t)$,\hfill(41b)
\end{flushleft}
\begin{flushleft}
\hspace{0.5in} $C_1 (t) = {\rm A}_1^1 \exp ( i(\omega _2 +
\omega_1) t) $\end{flushleft}
\begin{flushleft}
\hspace{0.5in} $ + {\rm A}_1^2 \exp (i( (\omega _2 + \omega_1) +
\Delta )t) + {\rm A}_1^3 \exp(i( (\omega _2 + \omega_1)  - \Delta
)t)$,\hfill(41c),
\end{flushleft}
where $A_i$-s are the constants which are calculated below from the
various initial conditions.
\par
Case-I: Let us consider initially at $t=0$, the atom is in level-1,
i.e, $C_1(0)=1$, $C_2 (0)=0$ and $C_3 (0)=0$. Using Eqs. (36) and
(41), the time dependent probabilities of the three levels are given
by
\begin{flushleft}
\hspace{1.5in} $\left| {C_3 (t)} \right|^2  = \frac{{\kappa _1^2
}}{{\Delta ^2 }}\sin ^2 \Delta t$,\hfill(42a)
\end{flushleft}
\begin{flushleft}
\hspace{1.5in} $\left| {C_2 (t)} \right|^2  = \frac{{\kappa _2^2
}}{{\Delta ^2 }}\sin ^2 \Delta t$,\hfill(42b)
\end{flushleft}
\begin{flushleft}
\hspace{1.5in} $\left| {C_1 (t)} \right|^2  = \cos ^2 \Delta
t$.\hfill(42c)
\end{flushleft}
\par
Case-II: If the atom is initially in level-2, i.e, $C_1 (0)=0$, $C_2
(0)=1$ and $C_3 (0)=0$, the corresponding probabilities of the
states are given by
\begin{flushleft}
\hspace{1.5in} $\left| {C_3 (t)} \right|^2  = 4\frac{{\kappa _1^2
\kappa _2^2 }}{{\Delta ^4 }}\sin ^4 \Delta t/2$,\hfill(43a)
\end{flushleft}
\begin{flushleft}
\hspace{1.5in} $\left| {C_2 (t)} \right|^2  = \frac{1}{{\Delta ^4
}}(\kappa _1^2 + \kappa _2^2 \cos \Delta t)^2$,\hfill(43b)
\end{flushleft}
\begin{flushleft}
\hspace{1.5in} $\left| {C_1 (t)} \right|^2  = \frac{{\kappa _2^2
}}{{\Delta ^2 }}\sin ^2 \Delta t$.\hfill(43c)
\end{flushleft}
\par
Case-III: When the atom is initially in level-3, i.e, $C_1 (0)=0$,
$C_2(0)=0$ and $C_3(0)=1$, we obtain the the occupation
probabilities of the three states as follows:
\begin{flushleft}
\hspace{1.5in} $\left| {C_3 (t)} \right|^2  = \frac{1}{{\Delta ^4
}}(\kappa _2^2 + \kappa _1^2 \cos \Delta t)^2$,\hfill(44a)
\end{flushleft}
\begin{flushleft}
\hspace{1.5in} $\left| {C_2 (t)} \right|^2  = 4\frac{{\kappa _1^2
\kappa _2^2 }}{{\Delta ^4 }}\sin ^4 \Delta t/2$,\hfill(44b)
\end{flushleft}
\begin{flushleft}
\hspace{1.5in} $\left| {C_1 (t)} \right|^2  = \frac{{\kappa _1^2
}}{{\Delta ^2 }}\sin ^2 \Delta t$.\hfill(44c)
\end{flushleft}
\vspace{0.5cm}
\begin{center}
\large VI.The quantized vee system
\end{center}
\par
The eigenfunction of the quantized vee system described by the
Hamiltonian in Eq.(10) is given by
\begin{flushleft}
\hspace{0.15in} $\left| {\Psi _{V} (t)} \right\rangle  =
\sum\limits_{n,m = 0}^\infty  {[C_1^{n +1,m} (t)\left| {n+1,m,1}
\right\rangle}+  C_2^{n,m} (t)\left| {n,m,2} \right\rangle +C_3^{n +
1,m - 1} (t)\left| {n +1,m -1,3} \right\rangle ].$
\end{flushleft}
\hspace {6.4in}(45)\\
Once again we note that the Hamiltonian couples the atom-field
states ${\left| {n+1,m,1} \right\rangle }$, ${\left| {n,m,2}
\right\rangle }$ and ${\left| {n + 1,m - 1,3} \right\rangle }$
forming vee configuration depicted in Fig.2. The interaction part of
the Hamiltonian (45) can also be expressed in the matrix form
\begin{flushleft}
\hspace{1.5in} ${\it H_{II}^{V}} = \hbar \left[
{\begin{array}{*{20}c}
   0 & 0 & {g_{1} \sqrt {m } }  \\
   0 & 0 & {g_{2} \sqrt {n + 1 } }  \\
   {g_{1} \sqrt {m } } & {g_{2} \sqrt {n + 1 } }& 0  \\
\end{array}} \right],$\hfill(46)
\end{flushleft}
and the corresponding eigenvalues are $\lambda _ \pm = \pm  \hbar
\sqrt {m g_{1}^2+(n+1)g_{2}^2}\quad (= \pm \hbar\Omega _{nm})$ and
$\lambda_0 = 0$ respectively. The dressed eigenstate is given by
\begin{flushleft}
\hspace{1.5in} $\left[ {\begin{array}{*{20}c}
   {\left| {nm, 3 } \right\rangle }  \\
   {\left| {nm,2} \right\rangle }  \\
   {\left| {nm, 1 } \right\rangle }  \\
\end{array}} \right] = T_{n,m}\left[ {\begin{array}{*{20}c}
   {\left| {n+1,m-1,3} \right\rangle }  \\
   {\left| {n,m,2} \right\rangle }  \\
   {\left| {n+1,m,1} \right\rangle }  \\
\end{array}} \right],$\hfill(47)
\end{flushleft}
the rotation matrix is found to be
\begin{flushleft}
\hspace{1.0in} $T_{n,m} = \left[ {\begin{array}{*{20}c}
   {g_{1} \sqrt {\frac{{m
}}{{2((n+1)g_{2}^2+ m g_{1}^2)}}}} & {g_{2} \sqrt {\frac{{n+1 }}{{2(
(n+1)g_{2}^2+ mg_{1}^2)}}}} & {\frac{1}{{\sqrt 2
}}}  \\
   {-g_{2} \sqrt {\frac{{n+1
}}{{(n+1)g_{2}^2+ m g_{1}^2}}}} & {g_{1} \sqrt {\frac{{m }}{{
(n+1)g_{2}^2 +
m g_{1}^2}}}} & {0}  \\
   {-g_{1} \sqrt {\frac{{m
}}{{2((n+1)g_{2}^2 + m g_{1}^2)}}}} & {-g_{2} \sqrt {\frac{{n+1
}}{{2((n+1)g_{2}^2+ m g_{1}^2 )}}}} & { \frac{1}{{\sqrt
2 }}}  \\
\end{array}} \right].$\hfill(48)
\end{flushleft}
The straightforward evaluation yields the various Euler angles are
\begin{flushleft}
\hspace{1.5in} $\theta_1=-\frac{\pi}{4}$, \quad
$\theta_2=\arccos[-\frac{\sqrt{n+1}g_2}{\sqrt{mg_1^2+(1+n) g_2^2}}
]$, $\theta_3=-\frac{\pi}{2}$. \hfill(49)
\end{flushleft}
The time-dependent probability amplitudes of the three levels are
given by
\begin{flushleft}
\hspace{.5in} $\left[ {\begin{array}{*{20}c}
   {C_3^{n + 1,m - 1} (t)}  \\
   {C_2^{n,m} (t)}  \\
   {C_1^{n + 1,m} (t)}  \\
\end{array}} \right] = {T^{-1}_{n,m}} \left[ {\begin{array}{*{20}c}
   {e^{-i\Omega _{nm} t} } & 0 & 0  \\
   0 & {e^{-i\Omega _{0} t}} & 0  \\
   0 & 0 & {e^{i\Omega _{nm} t} }  \\
\end{array}} \right]T_{n,m}\left[ {\begin{array}{*{20}c}
   {C_3^{n + 1,m - 1} (0)}  \\
   {C_2^{n,m} (0)}  \\
   {C_1^{n + 1,m} (0)}  \\
\end{array}} \right]$.\hfill(50)
\end{flushleft}
Once again we proceed to calculate the probabilities for different
initial conditions.
\par
Case-IV: Here we consider initially the atom is in level-1 i.e,
$C_1^{n +1,m}=1$, $C_2^{n,m}=0$ and $C_3^{n+1,m-1}=0$. Using
Eqs.(49)
 and
(50) the time-dependent  probabilities of the three levels are
given by
\begin{flushleft}
\hspace{1.5in} $\left| {C_3^{n+1,m-1} (t)} \right|^2  = \frac{{m
g_1^2}}{{\Omega _{nm}^2 }}\sin ^2 \Omega _{nm} t$,\hfill(51a)
\end{flushleft}
\begin{flushleft}
\hspace{1.5in} $\left| {C_2^{n,m} (t)} \right|^2  =
\frac{{(n+1)g_2^2 }}{{\Omega _{nm}^2 }}\sin ^2 \Omega _{nm}
t$,\hfill(51b)
\end{flushleft}
\begin{flushleft}
\hspace{1.5in} $\left| {C_1^{n + 1,m } (t)} \right|^2  = \cos ^2
\Omega _{nm} t$.\hfill(51c)
\end{flushleft}
Case-V: If the atom is initially in level-2 i.e, $C_3^{n+1,m-1}=0$,
$C_2^{n,m}=1$ and $C_1^{n+1,m}=0$, then corresponding probabilities
are
\begin{flushleft}
\hspace{1.5in} $\left| {C_3^{n + 1,m - 1} (t)} \right|^2  =
4\frac{{g_2^2 g_1^2 (n + 1 )(m )}}{{\Omega _{mn}^4 }}\sin ^4 \Omega
_{mn} t/2$,\hfill(52a)
\end{flushleft}
\begin{flushleft}
\hspace{1.5in} $\left| {C_2^{n,m} (t)} \right|^2  = \frac{1}{{\Omega
_{nm}^4 }}[m g_1^2+(n+1)g_2^2\cos \Omega _{nm} t]^2$,\hfill(52b)
\end{flushleft}
\begin{flushleft}
\hspace{1.5in} $\left| {C_1^{n+1,m} (t)} \right|^2  = \frac{{g_2^2
(n+1)}}{{\Omega _{nm}^2 }}\sin ^2 \Omega _{nm} t$.\hfill(52c)
\end{flushleft}
Case-VI: Finally if the atom is initially in level-3 i.e, $C_1^{n
+1,m}=0$, $C_2^{n,m}=0$ and $C_3^{n+1,m-1}=1$, then
\begin{flushleft}
\hspace{1.5in} $\left| {C_3^{n+1,m-1} (t)} \right|^2  =
\frac{1}{{\Omega _{nm}^4 }}[m g_1^2\cos \Omega _{nm}
t+(n+1)g_2^2]^2$,\hfill(53a)
\end{flushleft}
\begin{flushleft}
\hspace{1.5in} $\left| {C_2^{n,m} (t)} \right|^2  = 4\frac{{g_2^2
g_1^2 (n + 1 )(m )}}{{\Omega _{nm}^4 }}\sin ^4 \Omega _{nm}
t/2$,\hfill(53b)
\end{flushleft}
\begin{flushleft}
\hspace{1.5in} $\left| {C_1^{n + 1,m} (t)} \right|^2  = \frac{{m
g_1^2}}{{\Omega _{nm}^2 }}\sin ^2 \Omega _{nm} t$.\hfill(53c)
\end{flushleft}
Finally we note that for large values of n and m, Case-IV, V and VI
become identical to Case-I, II and III, respectively. This precisely
shows the validity of the Bohr's correspondence principle indicating
the consistency of our approach.
\begin{center}
\large VII.Numerical results and discussion
\end{center}
\par
Before going to show the numerical plots of the semiclassical and
quantized lambda and vee systems, we first consider their analytical
results. If we compare Case-I, II, III of both cases, we find that
the probabilities in Case-I (Case-III)) of lambda system is the same
as in Case-III (Case-I) of vee system except the populations of 1st
and 3rd levels are interchanged. See Eqs.(21 $\&$ 44) and Eqs.(23
$\&$ 42) for detailed comparison. Also Case-II respective models are
similar which is evident by comparing Eqs.(22 $\&$ 43). In contrast,
for the quantized model, Case-IV (Case-VI) of the lambda system is
no longer same as in Case-VI (Case-IV) of the vee system.
This breaking of symmetry is evident by comparing the analytical
results, Eqs.(31 $\&$ 53), Eqs.(32 $\&$ 52) and Eqs.(32 $\&$ 51)
respectively. Unlike previous case, also Case-V both the models are
different which is evident from Eqs.(22 $\&$ 43).
\par
In what follows, we compare the probabilities of the semiclassical
and quantized lambda and vee systems respectively. Fig.3 and 4 show
the plots of the probabilities $|C_{1}^i(t)|^2$ (blue line),
$|C_{2}^i(t)|^2$ (green line) and $|C_{3}^i(t)|^2$ (red line) for
the semiclassical lambda and vee models when the atom is initially
at level-1 (Case-I), level-2 (Case-II) and level-3 (Case-III),
respectively. The comparison of the plots shows that the pattern of
the probability oscillation of the lambda system for Case-I shown in
Fig.3a (Case-III in Fig.3c) is similar to that of Case-III shown in
Fig.4c (Case-I in Fig.4a) of the vee system. More particularly we
note that in all cases the oscillation of level-2 remains unchanged,
while the oscillation of level-3 (level-1) of the lambda system for
Case-I is identical to that of level-1 (level-3) of the vee system
for Case-III. Furthermore, comparison of Fig.3b and Fig.4b for
Case-II shows that the time evolution of the probabilities of
level-2 of both systems also remains similar while those of level-3
and level-1 are interchanged. From the behaviour of the probability
curve we can conclude that the lambda and vee configurations are
essentially identical to each other as we can obtain one
configuration from another simply by the inversion followed by the
interchange of probabilities.
\par
For the quantized field, we first consider the time evolution of the
probabilities taking the field is in a number state representation.
In the number state representation, the vacuum Rabi oscillation
corresponding to Case-IV, V and VI of the lambda and vee systems are
shown in Fig.5 and Fig.6 respectively. We note that, unlike
previous case, the Rabi oscillation for Case-IV shown in Fig.5a
(Case-VI shown in Fig.5c) for the lambda model is no longer similar
to Case-VI shown in Fig.6c (Case-IV shown in Fig.6a) for the vee
model. Furthermore, we note that for Case-V, the oscillation
patterns of Fig.5b is completely different from that of Fig.6b. In a
word, for the quantized field, in contrast to the semiclassical
case, the symmetry in the pattern of the vacuum Rabi oscillation in
all cases is completely spoiled irrespective of the fact whether the
system stays initially in any one of the three levels.
\par
The quantum origin of the breaking of the symmetric pattern of the
Rabi oscillation is the following. We note that due to the
appearance of the terms like $(n+1)$ or $(m+1)$, several elements in
the probabilities given by Eqs.(31,32,33) for the lambda system and
Eqs.(51,52,53) for the vee are non zero even at $m=0$ and $n=0$.
We argue that the vacuum Rabi oscillation interferes with the
probability oscillations of various levels and spoils their
symmetric structure. Thus as a consequence of the vacuum
fluctuation, the symmetry of probability amplitudes of the dressed
states of both models formed by the coherent superposition of the
bare states is also lost. In the other word, the invertibility
between the lambda and vee models exhibited for the classical field
disappears as the direct consequence of the quantization of the
cavity modes.
\par
Finally we consider the lambda and vee models interacting with the
bi-chromatic quantized fields which are in the coherent state. The
coherently averaged probabilities of level-3, level-2 and level-1
are given by
\begin{flushleft}
\hspace{1.5in} $\left\langle {P_3 (t)} \right\rangle _\Lambda   =
\sum\limits_{n,m} {W_n } W_m \left| {C_3^{n - 1,m } (t)} \right|^2$,
\hfill(54a)
\end{flushleft}
\begin{flushleft}
\hspace{1.5in} $\left\langle {P_2 (t)} \right\rangle _\Lambda   =
\sum\limits_{n,m} {W_n } W_m \left| {C_2^{n,m} (t)} \right|^2$,
\hfill(54b)
\end{flushleft}
\begin{flushleft}
\hspace{1.5in} $\left\langle {P_1 (t)} \right\rangle _\Lambda   =
\sum\limits_{n,m} {W_n } W_m \left| {C_1^{n - 1,m + 1} (t)}
\right|^2$, \hfill(54c)
\end{flushleft}
for the lambda system and
\begin{flushleft}
\hspace{1.5in} $\left\langle {P_3 (t)} \right\rangle _V   =
\sum\limits_{n,m} {W_n } W_m \left| {C_3^{n + 1,m - 1 } (t)}
\right|^2$, \hfill(55a)
\end{flushleft}
\begin{flushleft}
\hspace{1.5in} $\left\langle {P_2 (t)} \right\rangle _V   =
\sum\limits_{n,m} {W_n } W_m \left| {C_2^{n ,m } (t)} \right|^2$,
\hfill(55b)
\end{flushleft}
\begin{flushleft}
\hspace{1.5in} $\left\langle {P_1 (t)} \right\rangle _V   =
\sum\limits_{n,m} {W_n } W_m \left| {C_1^{n + 1,m } (t)} \right|^2$,
\hfill(55c)
\end{flushleft}
for the vee system, where $W_n=\frac{1}{n!}\exp[-\bar n]{\bar n}^n$
and $W_m=\frac{1}{m!}\exp[-\bar m]{\bar m}^m$ with $\bar n$ and
$\bar m$ be the mean photon numbers of the two quantized modes,
respectively. Fig.7-9 display the numerical plots of Eq.(54) and
Eq.(55) for Case-IV, V and VI respectively where the collapse and
revival of the Rabi oscillation is clearly evident for large average
photon numbers in both the fields. We note that in all cases, the
collapse and revival of level-2 of both the systems are identical to
each other. Furthermore, we note that the collapse and revival for
lambda system initially in level-1 shown in Fig.7a, Fig.7b and
Fig.7c (level-3 shown in Fig.9a, Fig.9b and Fig.9c) is the same as
that of the vee system if it is initially in level-3 shown in
Fig.7f, Fig.7e and Fig.7d (level-1 shown in Fig.9f, Fig.9e and
Fig.9d) respectively. On the other hand, if the system is initially
in level-2, the collapse and revival of the lambda systems shown in
Fig.8a, Fig.8b and Fig.8c are identical to Fig.8f, Fig.8e and Fig.8d
respectively for the vee system. This is precisely the situation
what we obtained in case of the semiclassical model. Thus the
symmetry broken in the case of the quantized model is restored back
again indicating that the coherent state with large average photon
number is very close to the classical state where the effect of
field population in the vacuum level is almost zero. It is needless
to say that the coherent state with very low average photon number
in the field modes can not show the symmetric dynamics in lambda and
vee systems.
\begin{center}
\large VIII.Conclusion
\end{center}
\par
This paper presents the explicit construction of the Hamiltonians of
the lambda, vee and cascade type of three-level configurations from
the Gell-Mann matrices of $SU(3)$ group and compares the exact
solutions of the first two models with different initial conditions.
It is shown that the Hamiltonians of different configurations of the
three-level systems are different. We emphasize that there is a
conceptual difference between our treatment and the existing
approach by Hioe and Eberly [18,21,22].
These authors advocate the existence of different energy conditions
which effectively leads to same cascade Hamiltonian ($h_{21}\neq0$,
$h_{32}\neq0$ and $h_{31}=0$ in Eq.(1)) having similar spectral
feature irrespective of the configuration. We justify our approach
by noting the fact that the two-photon condition and the equal
detuning condition is a natural outcome of our analysis. For the
lambda and vee models, the transition probabilities of the three
levels for different initial conditions are calculated while taking
the atom interacting with the bi-chromatic classical and quantized
field respectively. It is shown that due to the vacuum fluctuation,
the inversion symmetry exhibited by the semiclassical models is
completely destroyed. In other words, the dynamics for the
semiclassical lambda system can be completely obtained from the
knowledge of the vee system and vice versa while such recovery is
not possible if the field modes are quantized. The symmetry is
restored again when the field modes are in the coherent state with
large average photon number. Such breaking of the symmetric pattern
of the quantum Rabi oscillation is not observed in case of the
two-level Jaynes-Cummings model and therefore it is essentially a
nontrivial feature of the multi-level systems which is manifested if
the number of levels exceeds two. This investigation is a part of
our sequel studies of the symmetry breaking effect
for the equidistant cascade three-level and equidistant cascade
four-level systems respectively [36,37]. Following the scheme of
constructing of the model Hamiltonians, it is easy to show that we
have different eight dimensional Bloch equations and non-linear
constants for different configurations of the three-level systems
and these issues will be considered elsewhere [40]. The breaking of
the inversion symmetry of the lambda and vee models as a direct
effect of the field quantization is an intricate issue especially in
context with future cavity experiments with the multilevel systems.
\vfill
\begin{center}
\large Acknowledgement
\end{center}
MRN thanks University Grants Commission and SS thanks Department of
Science and Technology, New Delhi for partial financial support. We
thank Dr T K Dey for discussions. SS is also thankful to S N Bose
National Centre for Basic Sciences, Kolkata, for supporting his
visit to the centre through the Visiting Associateship program.
\pagebreak
\bibliographystyle{plain}

\pagebreak \vspace {2cm}
\begin{center}\begin{picture}(300,86)(0,0)

\Line(100,90)(200,90)\Text(210,89)[]{$\Delta_2$}
\Text(260,90)[]{$|3;m,n-1>$}

\ArrowLine(155,85)(170,40) \Text(175,70)[]{$\Omega_2$}
\Line(100,40)(200,40) \Text(250,40)[]{$|2;m,n>$}

\DashLine(100,80)(175,80){5} \DashLine(150,85)(200,85){5}

\LongArrow(110,100)(110,90) \LongArrow(110,70)(110,80)

\LongArrow(195,100)(195,90) \LongArrow(195,75)(195,85)

\ArrowLine(120,10)(150,80) \Text(140,20)[]{$\Omega_1$}

\Line(100,10)(200,10) \Text(265,10)[]{$|1;m+1,n-1>$}

\Text(90,85)[]{$\Delta_1$}

\end{picture} \\ {Fig.1 : Lambda type transition}
\end{center}
\vspace{1.5cm}
\begin{center}\begin{picture}(300,86)(0,0)

\Line(100,90)(200,90)

\Text(250,90)[]{$|3;m-1,n+1>$}

\ArrowLine(120,85)(140,10)

\Text(170,25)[]{$\Omega_2$}

\Line(100,40)(200,40)

\Text(250,40)[]{$|2;m,n>$}

\DashLine(100,85)(175,85){5}


\DashLine(140,35)(200,35){5}

\ArrowLine(145,10)(160,35)

\Text(120,60)[]{$\Omega_1$}

\Line(100,10)(200,10)

\Text(260,10)[]{$|1;m,n+1>$}


\LongArrow(110,100)(110,90) \LongArrow(110,78)(110,85)

\LongArrow(195,47)(195,40) \LongArrow(195,28)(195,35)

\Text(90,87)[]{$\Delta_1$}


\Text(210,39)[]{$\Delta_2$}

\end{picture} \\
{Fig.2 : Vee type transition}
\end{center}

\pagebreak
\begin{figure}[h]
\begin{center}
\rotatebox{0} {\includegraphics [width=16cm]{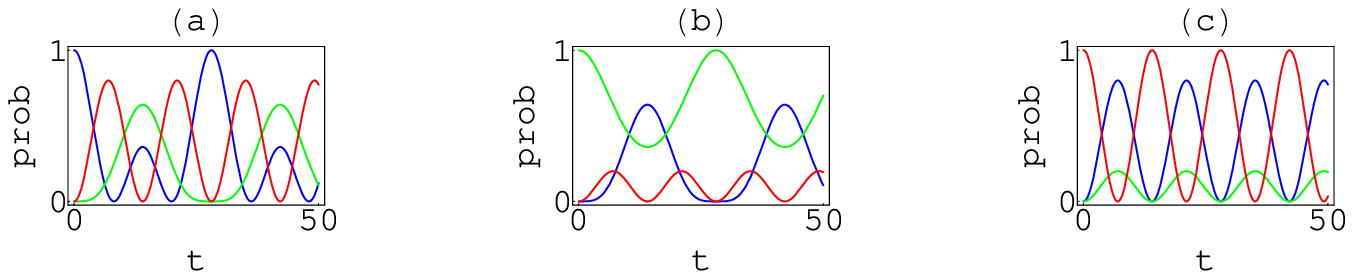}}
\end{center}
\noindent {\small {\bf [Fig.3]: The time evolution of the
probabilities $|C_{1}(t)|^2$ (blue line), $|C_{2}(t)|^2$ (green
line) and $|C_{3}(t)|^2$ (red line) of the semiclassical lambda
system for Case-I, II and III respectively with values ${\bf
\kappa_1=.2}$, ${\bf \kappa_2=.1}$. }}
\end{figure}

\begin{figure}[h]
\begin{center}
\rotatebox{0} {\includegraphics [width=16cm]{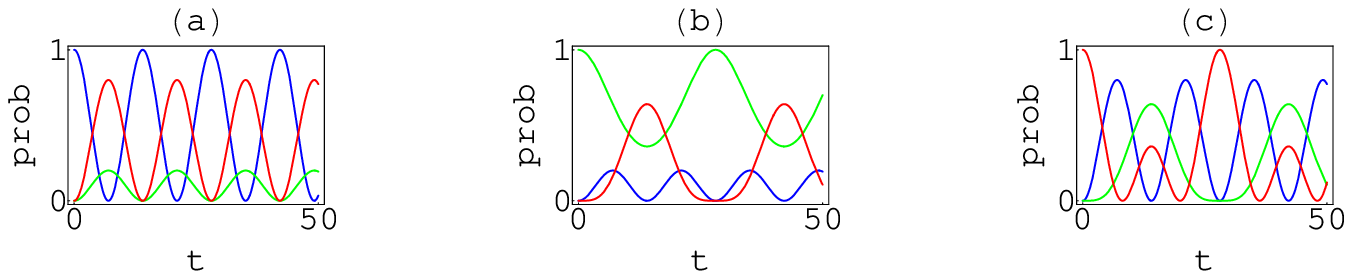}}
\end{center}
\noindent {\small {\bf [Fig.4]: The time variation of the
probabilities $|C_{1}(t)|^2$ (blue line), $|C_{2}(t)|^2$ (green
line) and $|C_{3}(t)|^2$ (red line) of the semiclassical vee system
for Case-I, II and III respectively with above values of ${\bf
\kappa_1}$, ${\bf \kappa_2}$.}}
\end{figure}

\pagebreak

\begin{figure}[h]
\begin{center}
\rotatebox{0} {\includegraphics [width=16cm]{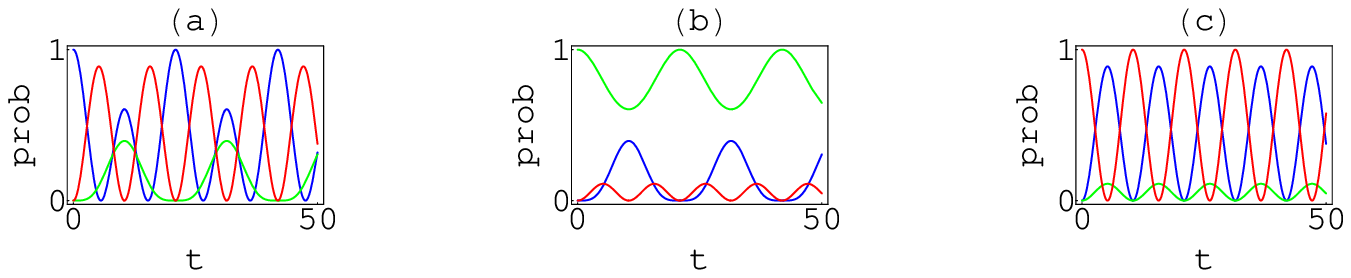}}
\end{center}
\noindent {\small {\bf [Fig.5]: The Rabi oscillation of the
quantized lambda system when the fields are in the number state for
Case-I, II and III, respectively with ${\bf g_1=.2}$, ${\bf
g_2=.1}$, ${\bf n=1}$, ${\bf m=1}$.}}
\end{figure}

\begin{figure}[h]
\begin{center}
\rotatebox{0} {\includegraphics [width=16cm]{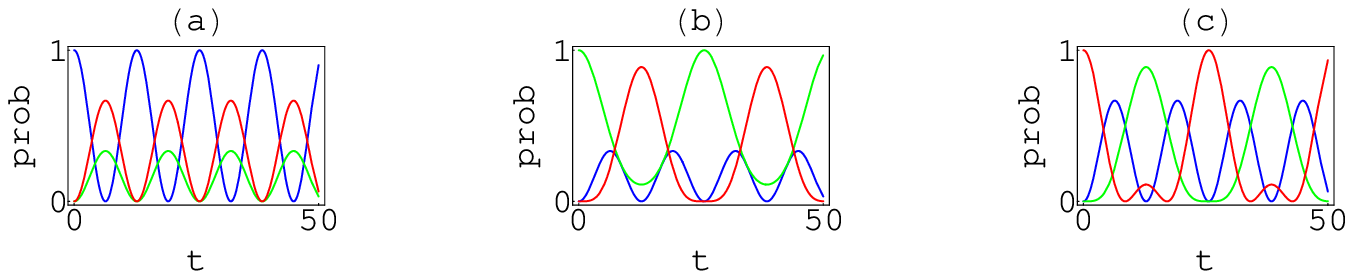}}
\end{center}
\noindent {\small {\bf [Fig.6]: The Rabi oscillation of the
quantized vee system when the fields are in the number states for
Case-I, II and III, respectively for same values of $\bf g_1$, $ \bf
g_2$, $\bf n$, $\bf m$.}}
\end{figure}

\pagebreak

\begin{figure}[h]
\begin{center}
\rotatebox{0} {\includegraphics [width=16cm]{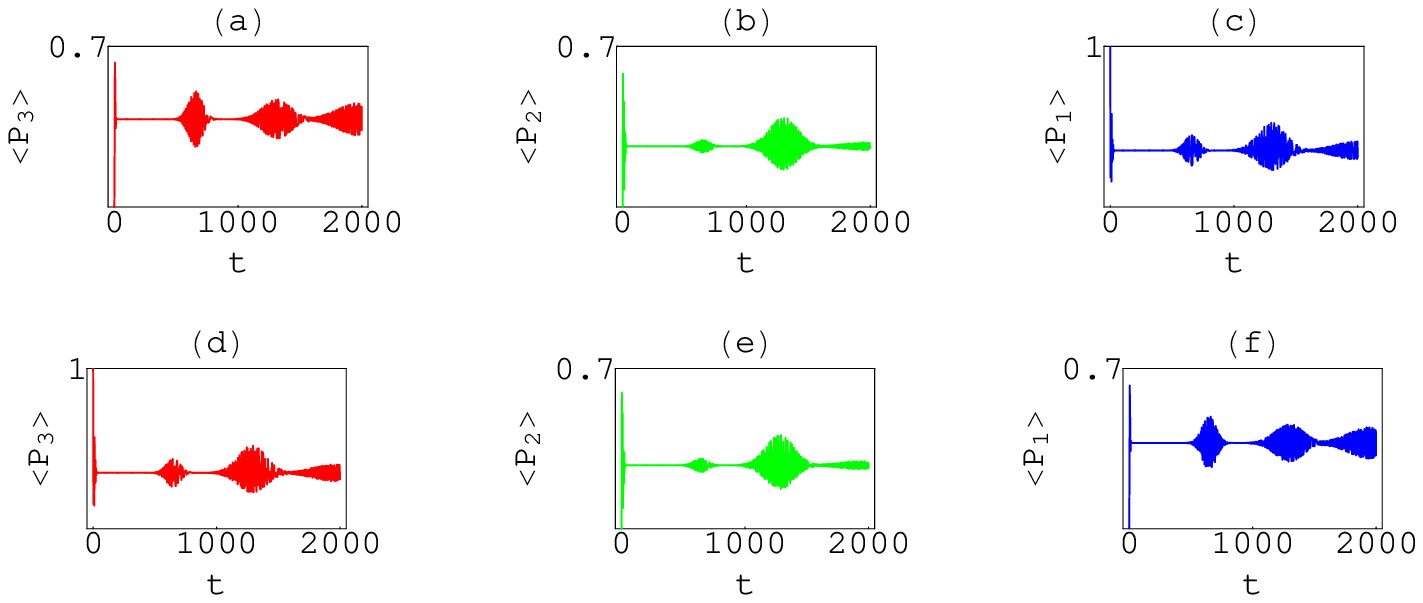}}
\end{center}
\noindent {\small {\bf [Fig.7]: Figs.7a-c display the time-dependent
collapse and revival phenomenon of level-3, level-2 and level-1 of
the lambda system for Case-IV, while Figs.7d-f show that of the
level-3, level-2 and level-1 respectively for of Case-VI of the vee
system taking the field modes are in coherent states with $\bar
n=30$ and $\bar m=20$.}}

\end{figure}

\pagebreak

\begin{figure}[h]
\begin{center}
\rotatebox{0} {\includegraphics [width=16cm]{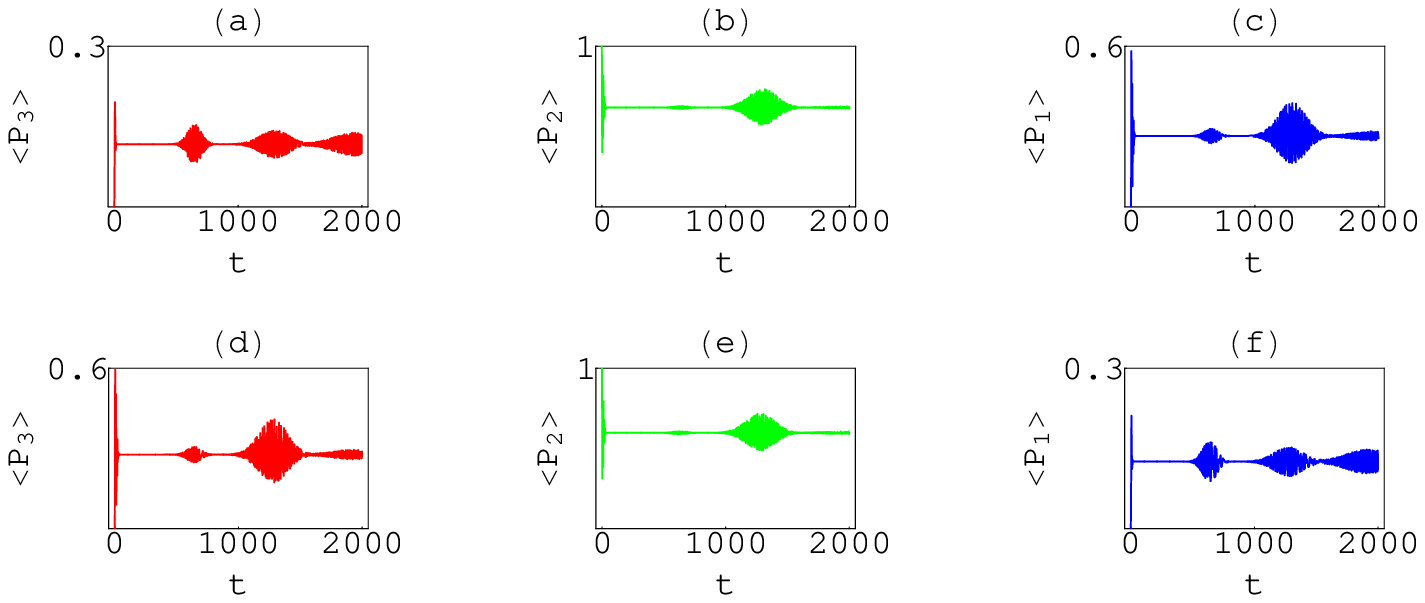}}
\end{center}
\noindent {\small {\bf [Fig.8]: Figs.8a-c display the time-dependent
of collapse and revival of level-3, level-2 and level-1 of the
lambda system for Case-V while Figs.8d-f show that of level-3,
level-2 and level-1 of the vee system for Case-V with the same
values of $\bar n$ and $\bar m$. as in Fig.7}}
\end{figure}

\pagebreak

\begin{figure}[h]
\begin{center}
\rotatebox{0} {\includegraphics [width=16cm]{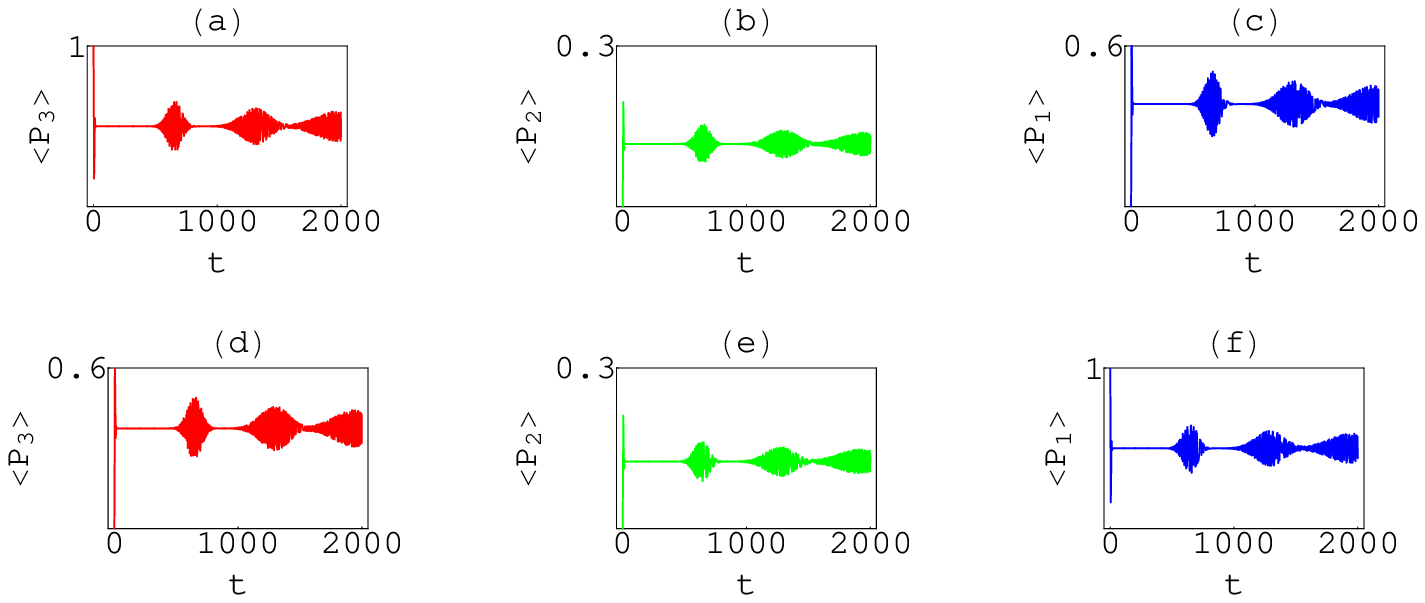}}
\end{center}
\noindent {\small {\bf [Fig.9]: Figs.9a-c display the time-dependent
of collapse and revival of level-3, level-2 and level-1 of the
lambda system for Case-VI while Figs.9d-f show that for level-3,
level-2 and level-1 respectively for the vee system for Case-IV with
the same values of $\bar n$ and $\bar m$ as in Fig.7.}}
\end{figure}

\end{document}